\documentclass{sig-alternate}

\usepackage{enumerate}
\usepackage{algorithm}
\usepackage{algorithmic}
\usepackage{array}
\usepackage{tabularx}
\usepackage{multirow}
\usepackage{subfigure}

\newcommand{\bc}{}
\newcommand{\md}{\mathbb{R}} 
\newcommand{\upT}{^\mathsf{T}}

\newcommand{\minv}{^{-1}}

\newcolumntype{I}{!{\vrule width 3pt}}
\newlength\savedwidth

\newlength\savewidth
\newcommand\shline{\noalign{\global\savewidth\arrayrulewidth
                            \global\arrayrulewidth 1.5pt}%
                   \hline
                   \noalign{\global\arrayrulewidth\savewidth}}

\begin{document}

\title{Analyzing User Preference for Social Image Recommendation}
%
%
%
%
%

\numberofauthors{3} 
%
%
%

\author{
\alignauthor
Xian-Ming Liu\\
       \affaddr{University of Illinois at Urbana-Champaign}\\
       \affaddr{Urbana, IL, USA}\\
       \email{xliu102@illinois.edu}\\
\alignauthor
Min-Hsuan Tsai\\
       \affaddr{Google Inc.}\\
\alignauthor
Thomas S. Huang\\
       \affaddr{University of Illinois at Urbana-Champaign}\\
       \affaddr{Urbana, IL, USA}\\
       \email{huang@ifp.uiuc.edu}\\
}


\maketitle

\begin{abstract}


With the incredibly growing amount of multimedia data shared on the social media platforms, recommender systems have become an important necessity
to ease users' burden on the information overload. In such a scenario, extensive amount of heterogeneous information such as tags, image content,
in addition to the user-to-item preferences, is extremely valuable for making effective recommendations. 
In this paper, we explore a novel hybrid algorithm termed {\em STM}, for image recommendation. STM jointly considers the problem of image content analysis with the users' preferences on the basis of sparse representation. 
STM is able to tackle the challenges of highly sparse user feedbacks and cold-start problmes in the social network scenario. 
In addition, our model is based on the classical probabilistic matrix factorization and can be easily extended to incorporate other useful information such as the social relationships. 
We evaluate our approach with a newly collected 0.3 million social image data set from Flickr. The experimental results demonstrate that sparse topic modeling of the image content leads to more effective recommendations, , with a significant performance gain over the state-of-the-art alternatives.

\end{abstract} 




\section{Introduction}
With the advent of social media web sites, such as {\em Facebook}, {\em Twitter}, {\em Flickr} and {\em Youtube},
the repositories of multimedia data shared is tremendous and unprecedented.
To ease the information overload, we targets at pursuing personalized recommendation algorithm which can analyze and understand the users' preferences
and
help users identify the information of their interest.

To facilitate the personalized recommendation, traditional methods tried to adopt the success from content-based information retrieval algorithms.
It performs recommendation by searching items with similar multimedia contents as the ones of users' interests.
Methods belong to this form is called Content-Analysis ({\em CA}) Recommender System, such as \cite{Yu:2009cw}, \cite{Yu:2009bm}.
CA relies on the content representation accuracy and coverage, such as image features, tags, and tries discover series of attributes to represent user' preferences.

However, it is suggested that the user behaviors, i.e., the ``human signal''
(e.g., comments, numeric ratings, binary ratings - ``Like'' on Facebook, and retwittes on Twitter)
tell us more about the content of multimedia than the content-based ones, especially for the web-scale data \cite{Slaney:2011ht}.
The human signal, known as collective information, can be efficient in bridging the semantic gap caused by content-based recommendation methods.

To utilize the collective information from human, {\em collaborative filtering} (CF) technique is proposed and has been extremely popular in
the study of recommender systems in the past decade. In the contrast of CA, it makes use of crowd wisdom and preferences, in addition to
personal interest, to establish recommendations for target users. In particular, a large amount of CF techniques adopted latent factors to characterize a user's or an item's profile with several unknown factors that learnt
from crowd's tastes toward the collection of items.

Although this type of CF methods gained the
state-of-the-art recommendation performance, they have several known limitations.
The latent factor based CF method can not handle cold start scenarios, where the
the items or users are unseen during the training stage. Such scenarios are critical to real-world
recommender systems, especially to the social media where new items are uploaded and shared at a
very high frequency.

Moreover, the sparsity of user feedbacks poses a great challenge on Collaborative Filtering based algorithms.
The Collaborative Filtering algorithms, especially matrix factorization methods,
rely on factorizing the user rating matrix into two latent distributions to represent user / item profiles.
However, when the rating matrix becomes sparse, the factorization could be extremely non-robust and easily destroyed by noise.
Compared with the movie or music recommendation in which the number of items is limited and the users are providing intensive feedbacks,
the social images are of huge amount and highly sparsely rated by users.
The average ``rating'' count for each Flickr image is $16.29$ according our statistics on $300,000$ photos crawled from 140 user groups (details in Section \ref{sec:dataset}).

Despite tentative efforts conducted on social image recommendation, the challenges are still unsolved.
Previous researches on social multimedia recommendation focus on recommending ``{\em communities}'' (e.g., Users Groups on Flickr)
instead of recommending items \cite{Chen:2008gu} \cite{Wang:2012gi} \cite{Yu:2009bm},
to avoid the sparsity problem existed in the rating matrix.
The community organization, which aggregates information from similar users or images,
increase the information density to facilitate efficient factorization.
However, the ``community'' methods only capture the common preferences, while lose the personalizations.

To tackle aforementioned challenges in the scenario of image recommendation in social networks,
we propose a novel hybrid framework in this paper
termed {\em Sparse Topic Models} (STM).
Motivated by the success of combining Content Analysis and Collaborative Filtering \cite{Melville:2002wv}\cite{Chong11Collaborative} in recommender system,
it incorporates both the content analysis and collaborative filtering in a uniformed optimization framework.
{\em STM} learns the user preference ({\em User Profiles}) and image representation ({\em Item Profiles}) from a joint user-image preference to link content analysis and collaborative filtering.
At the same time, sparse representation is introduced to handle the problem of sparsity existed in user rating matrices,
which makes the factorization more efficient and robust in social network settings.

More specifically, we involve the images' visual content into the latent-factor based CF (collaborative filtering) strategy
such that the factors in the latent spaces are specified by a {\em factor dictionary} which is associated with the sparse
visual bases and is used to construct the user profiles and image representation.
With the factor dictionary, a given specific factor in the latent space can be explained as visual patterns spanned on some sparse visual bases.

The intuitions behind the proposed method are three-fold.
Firstly, in spite of the importance of collective information, the item content could be equally critical in recommendation, especially for images.
People are usually aware of the visual patterns they prefer, and accordingly the users would be very likely to favor the images of similar appearance or patterns to
those images they already liked.
Secondly, we impose a sparse constraint when representing images in terms of the factor dictionary. This is a natural
and popular way in the computer vision community since an image usually represents a certain amount of patterns or factors
rather than ``thousands of words''. On the other hand, we have quantitatively identified an interesting and similar sparse
behavior on the user representation that a typical user would also not be interested in all kinds of patterns but focus
on only a few of them.
And finally, it allows on-the-fly encoding and construction of image profiles for ever growing social images,
thus has the ability to handle the cold start item problem by assuming the visually similar items reflect similar topics, which is shown in our experiments.

The proposed STM is of also merits in its generalization capability.
We further extend the model into {\em STM with Social Hints} (termed SoSTM) by involving the social hints between users to emphasize the social behaviors.
{\em SoSTM} utilizes the structures of users on social networks, to better characterize user preferences. It is verified both in \cite{ma08} and by our experiments that the social hints can help in convergence and improves the accuracy of recommendation.

Moreover, a real-world social image data set is collected
from Flickr, the {\em FlickrUserFavor} dataset,
which contains $350,000$ images, $20,000$ users, the users' favor toward images and social relationships between users. To the best of our knowledge, this is the first dataset for social image recommendation.
We targets at releasing this dataset along with the paper to share between the multimedia community.

The remainder of this paper is organized as follows. In Section \ref{sec:related}, we review some existing
work on collaborative filtering based recommender systems in the literature.
Section~\ref{sec:dataset} introduce the {\em FlickrUserFavor} dataset, the \emph{FlickrUserFavor}.
Then Section \ref{sec:stm} illustrate our proposed sparse topic model for recommendation, while an extend variation with the hints from social relationships is introduced in Section \ref{sec:stm_social}.
In Section \ref{sec:experiment} we present experiments on real-world data sets and show the advantages of
the proposed algorithm. The conclusions are presented in Section \ref{sec:conclusion}.

\section{Related Work}
\label{sec:related}

The main task of recommendation system is to predict the users' preferences on specific items according to historical ratings and feedbacks. Currently, the most popular approach is collaborative filtering (CF), which analyzes relationships between users and interdependencies among items to associate the new item and users \cite{dror2012web}, relying on previous user ratings. CF methods have been widely used in the Netflix Contest \cite{bennett2007netflix} and achieved fairly good performance \cite{koren08}.

Latent factor based, also known as model based methods are the most popular and effective CF methods where the users and the items are embedded in a latent space. In the latent space, each user or item is characterized as a vector which is called {\em profile}. The similarity of a user profile and a item profile in the latent space decides the user's preference toward the item.
There have been serval latent factor based CF methods studied in the literature, for example, \cite{Hofmann04,Sal07RBM,pmf08,koren2009matrix, salakhutdinov2008bayesian, agarwal2009regression}. In this paper, we will particularly focus on the {\em matrix factorization} methods \cite{pmf08,koren2009matrix}, which produce good predictive accuracy with great scalability. The basic principle lies behind the matrix factorization is to predict user rating of $i$-th user on $j$-th item ($r_{i,j}$) as:
\begin{equation}
    \hat{r}_{ij} = u_i^T v_j
\end{equation}
where $u$ is the user profile (subscript $i$ indicate $i$-th user), and $v$ is the item profile (again, subscript $j$ indicate $j$-th item).
These profiles are usually learned by minimizing the squared errors on the observed training data:
\begin{equation}
    \min_{u_*,v_*} \sum_{i,j}I_{ij}(r_{ij} - \hat{r}_{ij})^2 + \lambda(\|u_i\|^2 + \|v_j\|^2)
\label{eqn:mf}
\end{equation}
where $I_{ij}$ indicates whether the rating from user $i$ to item $j$ is observed ($I_{ij}=1$) or not ($I_{ij}=0$).
Various approaches have also been proposed to get more accurate predicted ratings, including SVD++ and its variations \cite{dror2012web, koren08} which approximates the ratings with a bias term ($b_{ij}$) and implicit feedbacks ($f(i)$).
\begin{equation}
    \hat{r}_{ij} = b_{ij}+ v_j^T(u_i + f(i))
\label{eqn:svdpp}
\end{equation}
%

Further researches consider the recommendation problem in the scenario of social network, by incorporating the social links between users into Eqn.~\eqref{eqn:mf} \cite{ma08,jamali10, ctr_smf12}. These approaches are referred to as \emph{social matrix factorization}. The idea here is to construct the user profile according to his or her social relationship (such as friendships) since closer social relationships may indicates similar tastes towards the items.

Another extension of collaborative filtering is to treat the latent space that user and item profiles lie in as a topic space following the basic idea of \emph{topic models} \cite{hofmann1999plsa, blei2003lda, barnard2003matching, blei2003modeling}. In such a situation, the user profile is a distribution of the user's preference over latent topics, and so is the item profile. Wang \emph{et al.} propose  CTR (collaborative topic regression) \cite{Chong11Collaborative} that leverages LDA (Latent Dirichlet Allocation) \cite{blei2003lda} to characterize both user profile and the document correlations. It assumes the user would prefer the documents of the similar latent topics as the ones he or she has rated before. It achieves fair improvement on scientific articles database.
%
%
Although CTR carries out an attempt to involve the document content in recommendation, it still remains unclear for the images due to the conclusion from previous study \cite{barnard2003matching} that topic model does not work well for visual features. The main difference between our work in this paper and the previous researches is, we derive a novel sparse topic model for collaborative filtering, which captures the image content and their correlations well, and is further utilized to social image recommendation.



\section{Social Image Recommendation \\ Data set}
\label{sec:dataset}


\subsection{Introduction to FlickrUserFavor Dataset}
The {\em FlickrUserFavor} data set is a social media data set, which is designed for the researches on social image recommendation, containing three types of information from the social domain and the multimedia domain.
The first type of information is the preference information that users indicate their favored images. The second and third type of information is the social information and the multimedia content information.
The key elements of the social information on {\em FlickrUserFavor} are the users and the groups. The users are those who upload, comment, tag, favor the images on {\em Flickr}. The groups, on the other hand, are communities with people who have the same interests toward a target subject. The group members typically favor photos which are closely related to the target subject. Therefore, we use this membership relation for user similarities where, in particular, users in the same group are similar to each other. The multimedia document on {\em Flickr} includes the images as well as the text or tags attached to each image.

\subsection{Dataset Composition}

350,000 images are collected from Flickr, which come from 140 user groups and are uploaded by 20,298 users. We use the {\em ``like''} feedback provided by users as a binary rating. 
%
For the tags, there are also more than 1,470,000 unique tags associated with these images, and we perform the TF-IDF to remove the stop words, and build a tag dictionary sized $5,000$ in our dataset.

One challenge with a prediction algorithm on such a dataset is that the ratings are extremely {\em sparse} (the sparsity of the user rating is $0.0008025$), and therefore the absence of a {\em ``like''} does not necessarily imply that the user does not like the image. 
Actually, the phenomenon behind is that users are tend to rate only a few images while each image is only rated by limited users.
This poses great challenge to traditional collaborative filtering methods.
Table~\ref{tab:dataset} shows the statistical details of the {\em FlickrUserFavor} dataset:

\begin{table}[h]
\centering
\caption{Statistical Details of {\em FlickrUserFavor} dataset.}
\label{tab:dataset}
\begin{tabular}{l|c}
\shline
Image Number                        &   $350,000$\\
\hline
User Number                         &   $20,298$\\
\hline
Tag Dictionary                      &   $5,000$\\
\hline
Average number of ratings per image &   $16.29$\\
\hline
Preference Sparsity                 &   $0.0008025$\\
\hline
Dimension of Visual Features (HG\cite{zhou2009hierarchical})    & $42,496$\\
\shline
\end{tabular}
\end{table}


\subsubsection{Training and Testing Data}
To make the training and testing to be justified and balanced, we use $75\%$ of the rating matrix as training, and the other $25\%$ (right bottom quarter) of the rating matrix as testing.
This partition ensures that there are information for each user and each image in the training stage, while the testing is kept independent as much as possible.

\subsection{Features and Measurements}


For the image representation, we adopt the Hierarchical
Gaussianization (HG) \cite{zhou2009hierarchical}, where each image is represented
by a Gaussian mixture model based super-vector. Specifically, all
the images are first resized to maximum 240 $\times$ 240 and
segmented into squared patches with three different sizes: 16, 25
and 31, by a 6-pixel step size. The 128-d Histogram of Oriented
Gradients (HOG) feature is then extracted from each patch and
followed by a PCA dimension reduction to 80-d. To obtain the feature
characteristics of the image collection, we learn a Gaussian mixture
model (GMM) with 512 components to obtain the statistics of
the patches of an image by adapting the distribution of the
extracted HOG features from these image patches.
The final dimension of the image feature is 42,496.

\subsubsection{Similarity Measurements}
As an important factor of social networks, the user correlation is also considered in our {\em FlickrUserFavor} dataset. We measure the correlations between users based on the membership of users. In particular, the social-similarity between user $U_A$ and $U_B$ is as follows:

\begin{align}
sim(U_A,U_B) \propto \bigg|Grp(U_A)\bigcap Grp(U_B)\bigg|
\end{align}
where $Grp(U)$ is the groups that user $U$ belongs to. For the text
domain the similarity is defined based on the co-occurring
words of the texts. Thus, the similarity between the text nodes
$T_A$ and $T_B$ is:
\begin{align}
sim(T_A,T_B) = \frac{\langle\delta(T_A),\delta(T_B)\rangle}
{\|\delta(T_A)\| \|\delta(T_B)\|}
\end{align}
where $\delta(T)$ is the bag-of-word representation of $T$ based on
the $5,000$ dimensional text codebook. For the image domain, the similarity is
defined based on the visual content similarity of the images.


\subsubsection{Evaluation Measurement}

Unlike the Netflix challenge\cite{bennett2007netflix} or Yahoo! Music dataset \cite{dror2012web}, where the RMSE (root mean square error) is widely adopted, the situation in our scenario is different, because the user rating in {\em FlickrUserFavor} dataset is binary rather than 5-level scores. Thus, it is biased using the RMSE as the evaluation measure for {\em FlickrUserFavor} dataset.

Instead, to evaluate the performance, we use an averaged ranked order of all the rated images in the testing dataset for a specific user.
In more details, we rank {\em all} the images by decreasing order of the algorithm recommendation scores for a relevant user $U$. Among these ranked images, we determine those for which the user $U$ has exhibited a {\em ``like''} preference in the {\em test data}, and report the {\em average percentile score} (APS) of the ranked images which are indeed preferred by the user $U$. We note that the lower the APS, the better the algorithm, which means the user preferred images are ranked in top positions. Finally, the {\em mAPS} is reported with the mean of the APS scores for all target users.

\section{Sparse Topic Model for Image \\ Recommendation}
\label{sec:stm}
In this section we introduce a new sparse topic model for image recommendation, which
incorporate image content analysis into the efficient recommendation scheme.

One of the main issues that the latent factor based recommendation models have been challenged
is the models cannot provide any explanations behind the recommendations being made.

\subsection{Model formulation}
For the recommendation tasks, there are two similar and closely related problems.
Given a set of observed ratings $\mathcal{R}$ from a group of users $\mathcal{U}$ on
a collection of images $\mathcal{I}$, we aim to predict the unobserved rating that a user
may give to the image. Another similar problem is to find a few top of images that the user
may be interested in.

We denote $X$ as the visual features of all the images in $\mathcal{I}$ such that
each column of $X\in\md^{d\times M}$, $X_j\in\md^{d}$, is the $d$-dimensional visual feature vector of
the $j$-th image. Also, we denote $V\in\md^{K\times M}$ as the image profile in terms of the topic
dictionary $D$.
Similar to $X$, each column in $V$, $V_j\in\md^{K}$, is the $K$-dimensional vector
representing the image in terms of the topics. Note that the dictionary $D$ is a $d$-by-$K$
matrix, each column of which represents a visual topic $D_k$.

We further denote $R\in\md^{N\times M}$ as the rating matrix where each row of $R$,
$R_{i.}\in\md^{M}$, is the ratings that $i$-th user gives to images and each column
of $R$, $R_{.j}\in\md^{N}$, is the ratings the all users give to $j$-th image.
Note that the matrix is usually very sparse as the amount of ratings that each user
can observe and provide are usually very limited. Hence we follow the way in classical
CF method \cite{pmf08, koren08} and introduce an observation matrix $I^R\in\md^{N\times M}$ such that
the $(i,j)$'s entry of $I^R$ is 1 if $i$-th user provides his/her rating on $j$-th image,
otherwise it is 0. The users in $\mathcal{U}$ are represented with $U\in\md^{K\times N}$,
where each column, $U_i\in\md^{K}$, is the user profile indicating $i$-th user's topical
preferences.

\begin{figure}[t]
\begin{center}
\centerline{\includegraphics[width=0.95\columnwidth]{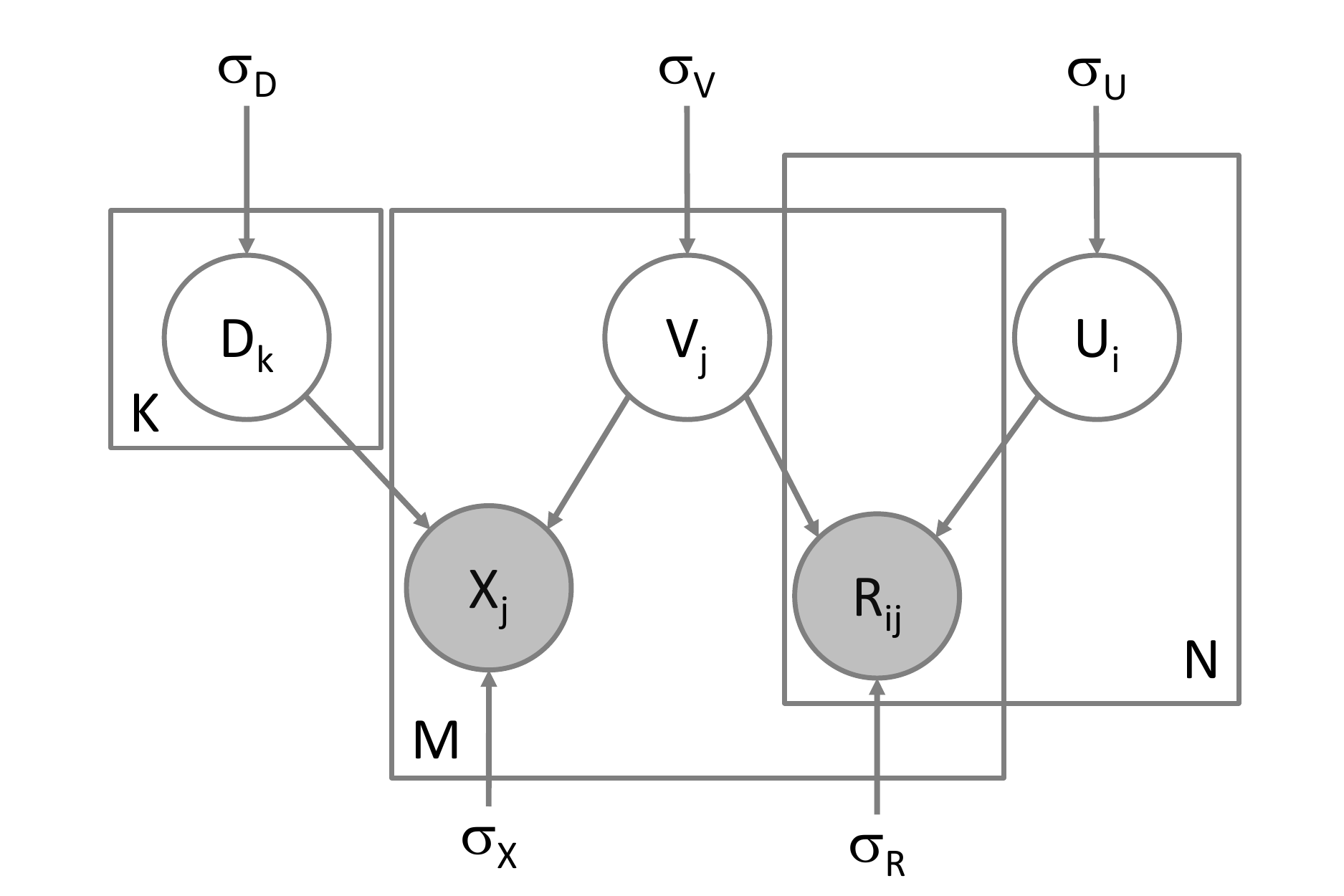}}
\caption{Graphical representation for sparse topic model}
\label{fig:stm_graphical_model}
\end{center}
\vskip -0.2in
\end{figure}

The observed image $X_j$ is assumed to be generated from a {\em topic dictionary} $D$ with $K$ topics,
so it can be represented on the bases of the topic dictionary, which we denoted as $V_j$.
The topical representation $V_j$ of the image is usually sparse since there are usually limited
topics within an image. The sparse topical representation in combine with the $i$-th user's
topical preferences, $U_i$, which is also sparse, generates the preference rating from the
$i$-th user to the $j$-th image. Since both the user and image profiles are builded upon the
sparse topics from the latent topic space, we name our model {\em sparse topic model}, or STM.
Figure \ref{fig:stm_graphical_model} illustrates the graphical model of STM.

We adopt a probabilistic linear model with Gaussian observation noise and define the conditional distribution over the
observed ratings as follows:
\begin{align}
p(X|D,V,\sigma^2) &= \prod_j  \mathcal{N}(X_{j}|DV_j,\sigma^2 \mathbf{I})  \\
p(R|U,V,\sigma_R^2) &= \prod_i\prod_j \left( \mathcal{N}(R_{ij}|U_i^TV_j,\sigma_R^2) \right)^{I_{ij}}
\label{eqn:cond_prob}
\end{align}
we also place zero-mean Laplacian priors on image and user profiles to enforce sparsity constraints on the user and image profiles:
\begin{align}
p(U|\sigma_U^2) &= \prod_i Laplace(U_i|0,\sigma_U^2 \mathbf{I}) \\
p(V|\sigma_V^2) &= \prod_j Laplace(V_j|0,\sigma_V^2 \mathbf{I})
\label{eqn:lap_prior}
\end{align}
Therefore, the log of the posterior distribution over the topic dictionary $D$, user profiles $U$ and image profiles $V$ is
\begin{align}
&\ln p(D,U,V|X,R,\sigma^2,\sigma_R^2,\sigma_U^2,\sigma_V^2) \propto \nonumber\\
&~~~~~~~~~-\frac{1}{2\sigma^2} \sum_j (X_j-DV_j)^T(X_j-DV_j) \nonumber\\
&~~~~~~~~~-\frac{1}{2\sigma_R^2} \sum_i\sum_j I^R_{ij}(R_{ij}-U_i^TV_j)^2 \nonumber\\
&~~~~~~~~~-\frac{1}{2\sigma_U^2} \sum_i \|U_i\|_1 -\frac{1}{2\sigma_V^2} \sum_j \|V_j\|_1
\end{align}
Therefore, maximizing a posteriori estimation of $D$, $V$, $U$ can be equivalently formulated as the following optimization problem:
\begin{align}
\min_{D,V,U} &~\mathcal{L} = \frac{1}{2}\|X-D V\|_F^2 + \frac{\lambda_R}{2}\|I^R\circ(R-U\upT V)\|_F^2 + \nonumber \\
&~~~~~~~~ \lambda_U \sum_i \|U_i\|_1 + \lambda_V \sum_j \|V_j\|_1 \nonumber\\
\text{s.t.} &~~ \|D_k\|_2^2 \le 1,  ~\forall k
\label{eqn:obj}
\end{align}
with $\lambda_R = \sigma^2/\sigma_R^2$, $\lambda_U = \sigma^2/\sigma_U^2$, $\lambda_V = \sigma^2/\sigma_V^2$.


\subsection{Learning Algorithm} \label{subsec:learning}
To learn the sparse topic model, $D,V,U$, in the optimization problem \eqref{eqn:obj},
we decompose the problem into three types of subproblems as follows:
\begin{align}
(SP_{D}) & \min_{D} \frac{1}{2}\|X-D V\|_F^2~~~ \text{s.t.} ~ \|D_k\|_2^2 \le 1,  ~\forall k \label{eqn:sp_d}\\
(SP_{V_j}) & \min_{V_j} \frac{1}{2}\|X_j-D V_j\|^2 + \frac{\lambda_R}{2}\|I^R_{.j}\circ(R_{.j}-U\upT V_j)\|^2 + \nonumber\\
&~~~~~ \lambda_V \|V_j\|_1 \label{eqn:sp_v_j} \\
(SP_{U_i}) & \min_{U_i} \frac{\lambda_R}{2}\|I^R_{i.}\circ(R_{i.}-U_i\upT V)\|^2 + \lambda_U \|U_i\|_1
\label{eqn:sp_u_i}
\end{align}

For the first subproblem ($SP_D$) it is not difficult to obtain an analytical solution with
the dual variables $\lambda_{D_k}$ for all $k$ constraints:
\begin{align}
D = (XV\upT)(\text{diag}(\lambda_{D_k}) + VV\upT)\minv
\label{eqn:sp_d_sol}
\end{align}
The dual variables $\lambda_{D_k}$ for $k$-th constraint can be obtained by maximizing the Lagrange dual of
subproblem ($SP_{D}$), for more details refer to \cite{lee06}.

For the second and third type of subproblems,
we employ the feature-sign search algorithm \cite{lee06} after arranging them in a canonical form as follows:
\begin{align}
(SP_{V_j}) & \min_{V_j} \frac{1}{2}V_j\upT P_{V_j} V_j - q_{V_j}\upT V_j + \lambda_V \|V_j\|_1
\label{eqn:sp_v_j_c} \\
(SP_{U_i}) & \min_{U_i} \frac{1}{2}U_i\upT P_{U_i} U_i - q_{U_i}\upT U_i + \lambda_U \|U_i\|_1
\label{eqn:sp_u_i_c}
\end{align}
where
\begin{align}
P_{V_j} &= D\upT D + \lambda_R \hat{U} \hat{U}\upT \\
q_{V_j} &= D\upT X_j + \lambda_R \hat{U}\hat{R}_{.j} \\
P_{U_i} &= \lambda_R \hat{V}\hat{V}\upT \\
q_{U_i} &= \lambda_R \hat{V}\hat{R}_{i.}\upT
\end{align}
Note that the hatted variables are a submatrix or subvector due to the observation matrix $I^R$.
For instance, $\hat{U} = U_{I_{.j}}$ indicates the submatrix of user profiles that correspond to
the users with observed ratings on $j$-th image.
With the canonical formulation, we can plug-in our subproblems to the feature-sign search algorithm with
the following casts ($P \equiv A\upT A,~q\equiv 2A\upT y$):
\begin{align}
\frac{\partial{\|y-Ax\|^2}}{\partial{x_i}} &= P_{i.}x - q_i \\
\hat{x}_{new} &= \frac{1}{2} P_{\Lambda\Lambda}\minv (q_\Lambda - \gamma\theta_\Lambda)
\end{align}
where $P_{i.}$ is the $i$-th row of matrix $P$ and $q_i$ is the $i$-th entry of the vector $q$.
Also, we use $\Lambda$ to denote the active set, and $P_{\Lambda\Lambda}$ is a submatrix of $P$
that contains only the columns and the rows corresponding to the active set.

%
%
%
%


\subsection{Dealing with Cold-Start Images}
The new image or the cold start image problem refers to making recommendations of unseen and unrated images to users.
Although it is very common and practical for the recommendation task,
the traditional matrix factorization methods \cite{pmf08,koren08} could not handle such a problem as
there is no way to obtain the item profiles in the testing time.
On the other hand, with the topic dictionary $D$ learned during the training time, our proposed model is able to
efficiently encode the cold start images during the testing time.

Given an unseen image $X_{new}$ without any available ratings from users, it image profile $V_{new}$
can be obtained through the following sparse encoding process:
\begin{align}
\min_{V_{new}} \frac{1}{2}\|X_{new}-D V_{new}\|^2 + \lambda_V \|V_{new}\|_1
\label{eqn:sp_new_image}
\end{align}
which can be solved efficiently using the technique similar to that stated in section \ref{subsec:learning} with the following coefficient matrices:
\begin{align}
P_{V_{new}} = D\upT D, ~~~ q_{V_{new}} = D\upT X_{new}
\end{align}

\section{Recommendation with Social \\ Hints}
\label{sec:stm_social}

\begin{figure}[t]
\begin{center}
\centerline{\includegraphics[width=1.05\columnwidth]{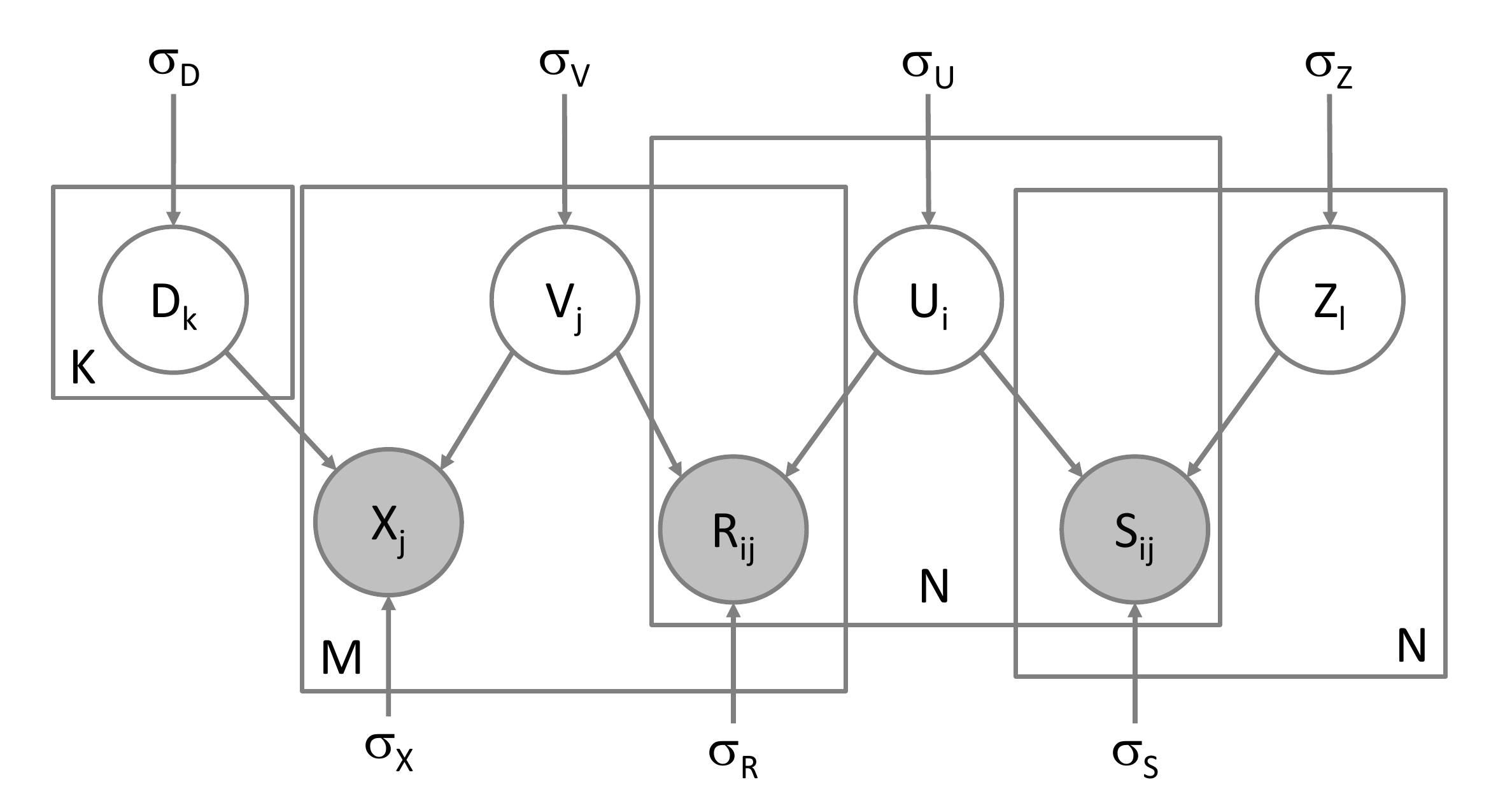}}
\caption{Graphical representation for sparse topic model with social hints}
\label{fig:stm_social}
\end{center}
\vskip -0.2in
\end{figure}

In this section, we will extend our STM approach with the incorporation of the social relationships.
We follow the idea from SoRec \cite{ma08} to introduce the {\em factor profile} for each user which
captures the topic preferences of the user based on his or her social relationships
instead of the user's own preference.
With the factor profiles, the social relationship between two users can be inferred from the affinity
between the user profile from one user and the factor profile from the other user.
Figure \ref{fig:stm_social} illustrates how the factor profiles $Z$ involve in generating the social relationships
$S$.

As indicated in Figure \ref{fig:stm_social},  we extend the sparse topic model with social relationships by jointly considering
the generation of social relationship as follows:
\begin{align}
\min_{D,V,U} &~\mathcal{L} = \frac{1}{2}\|X-D V\|_F^2 + \frac{\lambda_R}{2}\|I^R\circ(R-U\upT V)\|_F^2 + \nonumber \\
&~~~~~~~~ \frac{\lambda_S}{2}\|I^S\circ(S-U\upT Z)\|_F^2 + \lambda_U \sum_i \|U_i\|_1 +  \nonumber\\
&~~~~~~~~ \lambda_V \sum_j \|V_j\|_1 + \lambda_Z \|Z\|_F^2 \nonumber \\
\text{s.t.} &~~ \|D_k\|_2^2 \le 1,  ~\forall k
\label{eqn:stms_obj}
\end{align}
where $Z\in\md^{k\times N}$ is the factor profile matrix so that each of its column $Z_i\in\md^{k}$ represents the factor profile
for the $i$-th user. $S$ is a $N$-by-$N$ matrix whose $(i,j)$'s entry indicates the social relationship between $i$-th and $j$-th users.
The observation matrix $I^S\in\md^{N\times N}$ is defined that
the $(i,j)$'s entry of $I^S$ is 1 if $i$-th and $j$-th users have social relationship, otherwise it is 0.
Similarly, we decompose the problem into the following four subproblems:
\begin{align}
(SP_{D}) & \min_{D} \frac{1}{2}\|X-D V\|_F^2~~~ \text{s.t.} ~ \|D_k\|_2^2 \le 1,  ~\forall k \label{eqn:stms_sp_d}\\
(SP_{V_j}) & \min_{V_j} \frac{1}{2}\|X_j-D V_j\|^2 + \frac{\lambda_R}{2}\|I^R_{.j}\circ(R_{.j}-U\upT V_j)\|^2 + \nonumber\\
&~~~~~~~~ \lambda_V \|V_j\|_1 \label{eqn:stms_sp_v_j} \\
(SP_{U_i}) & \min_{U_i} \frac{\lambda_R}{2}\|I^R_{i.}\circ(R_{i.}-U_i\upT V)\|^2 +  \nonumber \\
&~~~~~~~~  \frac{\lambda_S}{2}\|I^S_{i.}\circ(S_{i.}-U_i\upT Z)\|^2 + \lambda_U \|U_i\|_1 \label{eqn:stms_sp_u_i} \\
(SP_{Z_i}) & \min_{Z_i} \frac{\lambda_S}{2}\|I_{.i}^S\circ(S_{.i}-U\upT Z_{i})\|^2 +  \lambda_Z \|Z_{i}\|^2 \label{eqn:stms_sp_z}
\end{align}
Note that we can again write the subproblem $(SP_{V_j})$ and $(SP_{U_i})$ with the canonical form as those in \eqref{eqn:sp_v_j_c}
and \eqref{eqn:sp_u_i_c} with the following coefficients:
\begin{align}
P_{V_j} &= D\upT D + \lambda_R \hat{U} \hat{U}\upT \\
q_{V_j} &= D\upT X_j + \lambda_R \hat{U}\hat{R}_{.j} \\
P_{U_i} &= \lambda_S \hat{Z}\hat{Z}\upT + \lambda_R \hat{V}\hat{V}\upT \\
q_{U_i} &= \lambda_S \hat{Z}\hat{S}_{i.}\upT + \lambda_R \hat{V}\hat{R}_{i.}\upT
\end{align}
Also, the subproblem $(SP_{Z_i})$ is simply a regularized unconstrained quadratic programming that has an analytic solution:
\begin{align}
Z_i = (\hat{U} \hat{U}\upT + \frac{2\lambda_Z}{\lambda_S}\mathbf{I})^{-1}\hat{U}\hat{S}_{.i}
\end{align}

\section{Experimental Results}
\label{sec:experiment}

In this section, we compare our proposed algorithm with the other state-of-the-art recommendation algorithms,
on the {\em FlickrUserFavor} data set.
Especially, we also conduct the experiments on cold-start images to testify the improvement of our algorithm by introducing the content factor and sparse coding in the collaborative filtering.

\subsection{Baseline Methods}
In our experiments, we mainly compare with several state-of-the-arts latent factor based CF methods, namely PMF \cite{pmf08}, SoRec \cite{ma08}, and CTR \cite{Chong11Collaborative}.

Probabilistic Matrix Factorization (PMF) \cite{pmf08}
is a probabilistic latent factor-based approach proposed for
collaborative filtering. PMF aims at fitting the user-item rating
matrix using low-rank approximations, that is, the $\ell$-dimensional
user and item factor matrices ($\ell$ is small). The method, however, ignores
the social activities and content information and simply assumes
the users and items are independent and identically distributed.
In this experiment, we use 30-dim latent features for both users and items.

SoRec \cite{ma08} is an extended version of PMF which fuses the user-item
rating matrix with the user's social network. The tuition behind
the approach that people in a user's social network affect the user's
personal behaviors more than those not in the user's network do.
In this experiment, we use 30-dim latent vectors for user-specific,
factor-specific (social relations) and item-specific feature vectors.


CTR \cite{Chong11Collaborative} is one of the most recent work on involving document content in the CF. It combines the merits of traditional collaborative filtering and topic modeling in a probabilistic model, by assuming the documents of similar contents should have similar item-profile.
Despite the positive, the usage of CTR is limited in our application because of two reasons. First, it has poor scalability, because the parameters to estimate boost when the data scale increases; On the other hand, unlike the words and sentences in articles, there are no natural boundaries between visual bases for images even though after image segmentation.

However, we found that our model can be modified slightly into a simplified {\em collaborative topic regression} model which is adapted for visual features, by relaxing the constraint of both user-profile and item-profile matrices $U$ and $V$ by $L2$ norm instead of $L1$, as:
\begin{align}
\min_{D,V,U} &~\mathcal{L} = \frac{1}{2}\|X-D V\|_F^2 + \frac{\lambda_R}{2}\|I\circ(R-U\upT V)\|_F^2 + \nonumber \\
&~~~~~~~~ \lambda_U \sum_i \|U_i\|_2^2 + \lambda_V \sum_j \|V_j\|_2^2 \nonumber\\
\text{s.t.} &~~ \|D_k\|_2^2 \le 1,  ~\forall k
\label{eqn:obj_ctr}
\end{align}

\begin{figure}[t]
\begin{center}
\centerline{\includegraphics[width=1.05\columnwidth]{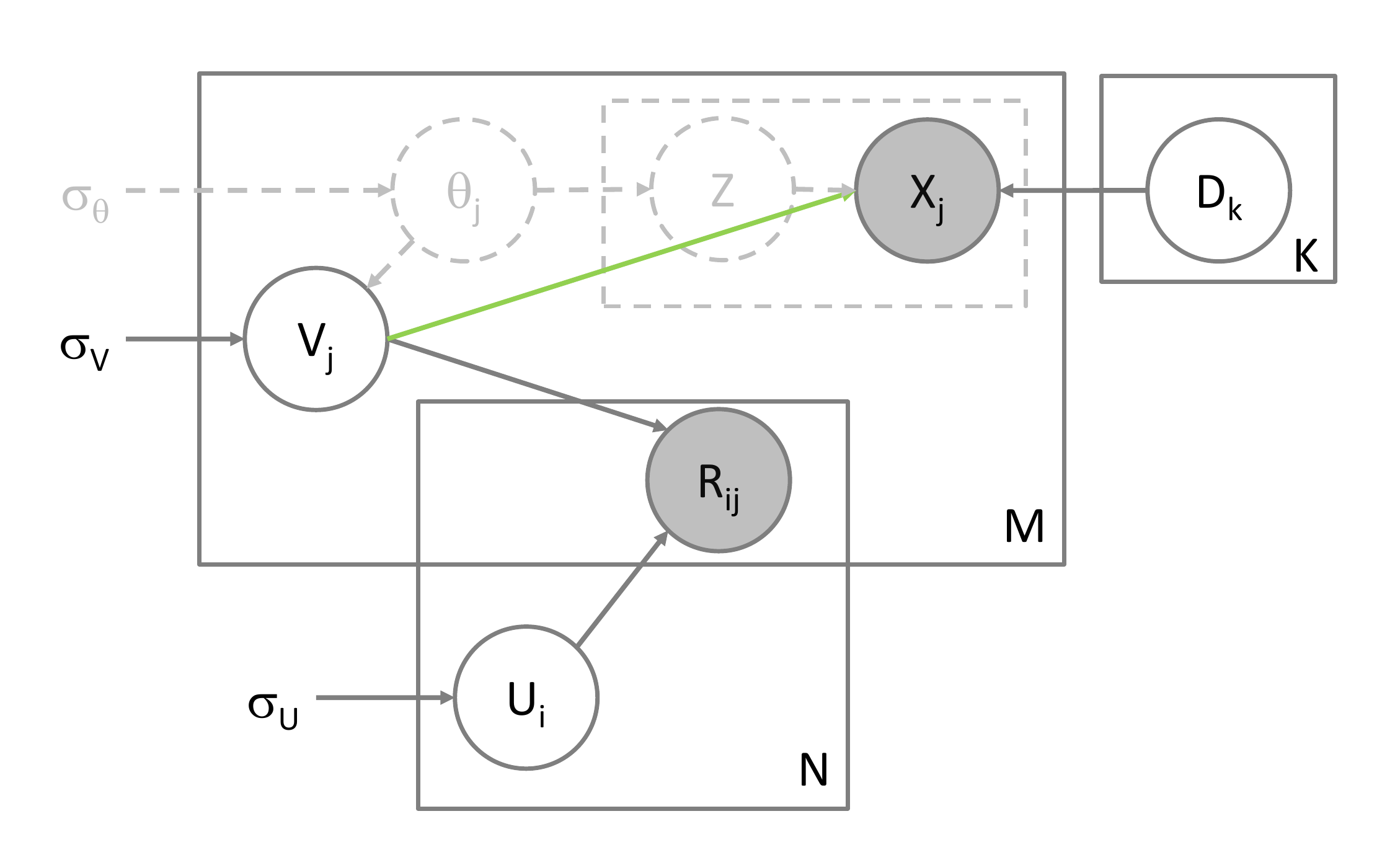}}
\caption{Graphical model of the original CTR model (with light grey dashed plates) and the modified CTR model for image (CTR-I, with green line)}
\label{fig:ctr_i_graph_model}
\end{center}
\end{figure}

Similarly, we can still decompose the optimization in Eqn~\eqref{eqn:obj_ctr} into two parts: the {\em collaborative filtering} subproblem
\begin{align*}
\min_{U,V} \frac{\lambda_R}{2}\|I\circ(R-U\upT V)\|_F^2 + \lambda_U \sum_i \|U_i\|_2^2 + \frac{\lambda_V}{2} \sum_j \|V_j\|_2^2
\end{align*}
and the {\em topic modeling} subproblem
\begin{align*}
\min_{D,V}\frac{1}{2}\|X-D V\|_F^2 + \frac{\lambda_V}{2} \sum_j \|V_j\|_2^2 ~~\text{s.t.} ~~\|D_k\|_2^2 \le 1,  ~\forall k
\end{align*}
We still use the iterative method similar with the techniques in section \ref{sec:stm} to solve the problem. For convenient, we name this as {\em CTR-I}.

The detailed differences among the methods are shown in Table~\ref{tab:methods}.

\begin{table}[h]
\centering
\caption{Comparison with various latent factor based approaches.}
\label{tab:methods}
\begin{tabular}{|c|c|c|c|}
\shline
                                    &   Social          & Content                       &   Scalability\\
\hline
PMF \cite{pmf08}                    &                   &                               &   Good\\
\hline
SoRec \cite{ma08}                   &   $\checkmark$    &                               &   Good\\
\hline
CTR \cite{Chong11Collaborative}     &                   & $\checkmark$                  &   Poor\\
\hline
\textbf{STM}                        &                   & $\checkmark$                  &   Good\\
\hline
\textbf{SoSTM}                      &   $\checkmark$    & $\checkmark$                  &   Good\\
\shline
\end{tabular}
\end{table}

\subsection{Algorithm Robustness}
\subsubsection{Convergence}

\begin{figure}[h]
\begin{center}
\centerline{\includegraphics[width=1.05\columnwidth]{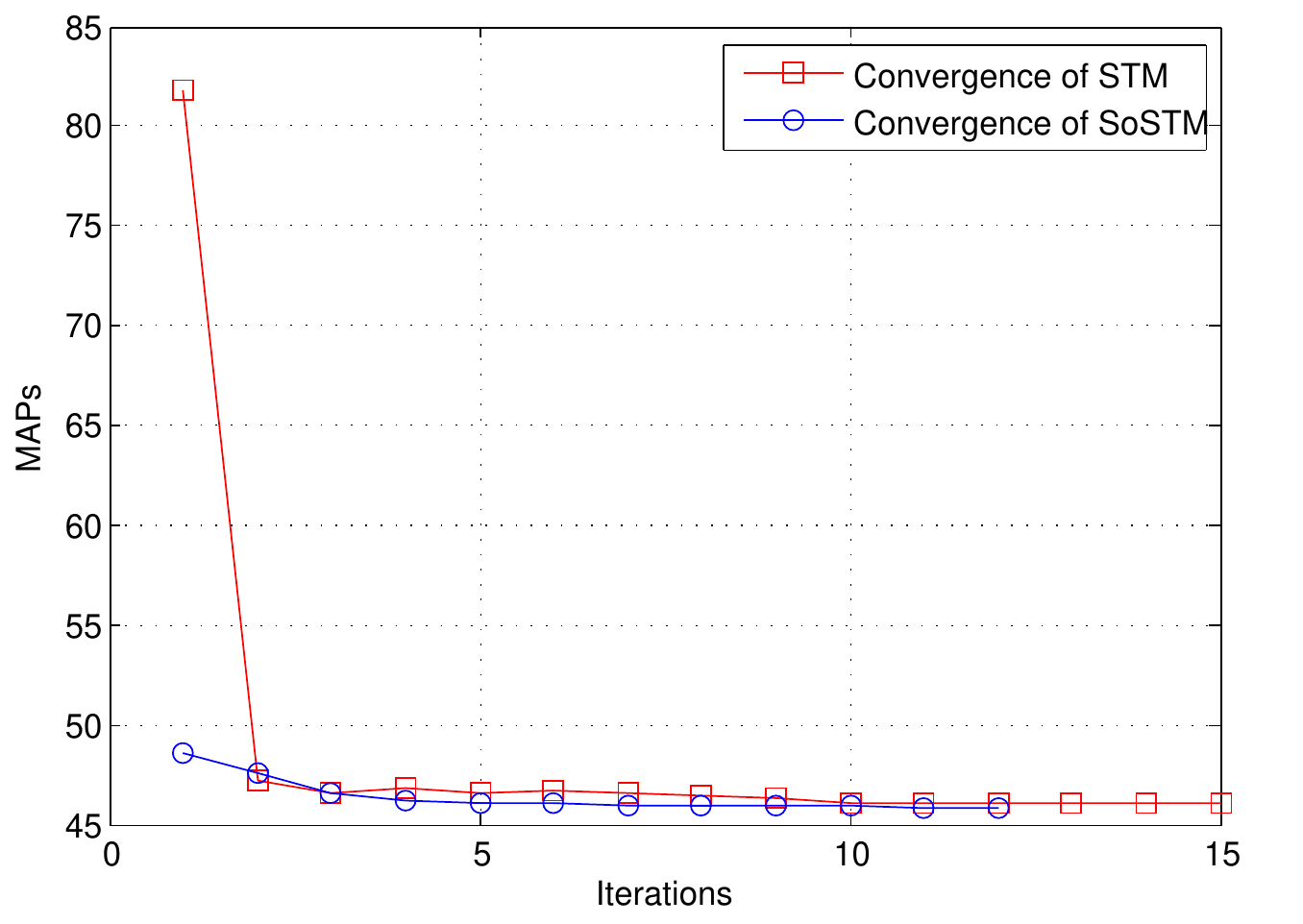}}
\caption{Convergence of Sparse Topic Modeling: the recommendation performance on different iterations.}
\label{fig:convergence}
\end{center}
\end{figure}

Figure~\ref{fig:convergence} shows the performance of \emph{STM} on different iterations. According to Figure~\ref{fig:convergence}, the algorithm will converge sharply after first two iterations, and achieves a good mAPS at 10-th iteration, which is $45.91$. Besides the theoretical analysis in Section~\ref{sec:stm}, this experimental evidence also guarantees the efficiency of our algorithm.

\vspace{0.1in}
\noindent
{\em How Social Hints improve Convergence?}\\
\indent Figure~\ref{fig:convergence} also shows the convergence of {\em SoSTM} algorithm proposed in Section~\ref{sec:stm_social}. As shown in the results, the {\em SoSTM} converges more smoothly than standard STM, benefit from the extra social hints involved in the training.

Furthermore, with the help of social hints, the mAPS in the first iteration improves from $81.79$ to $48.54$. Especially, the social hints make the initialization of user profile more stable, because the user preference are propagated to related users (in our scenario, the users who are in the same interest group on Flickr) even in the early stage of the algorithm.

\subsubsection{Parameter Validation}

\begin{figure*}[htb]
  \centering
  \subfigure[mAPS vs. $\lambda_U$]{
    \label{fig:lambda_U}
    \includegraphics[width=0.95\columnwidth]{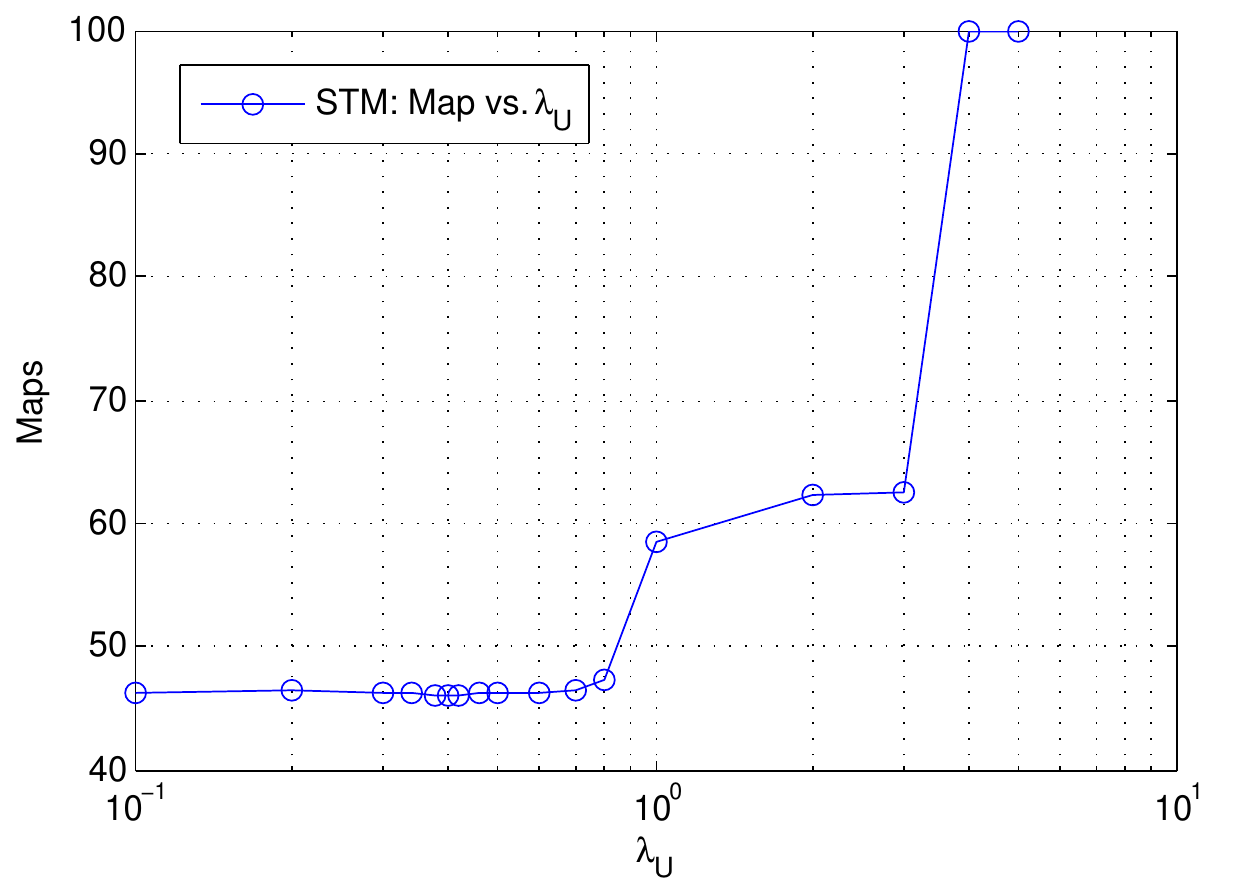}}
  \subfigure[mAPS vs. $\lambda_V$]{
    \label{fig:lambda_V}
    \includegraphics[width=0.95\columnwidth]{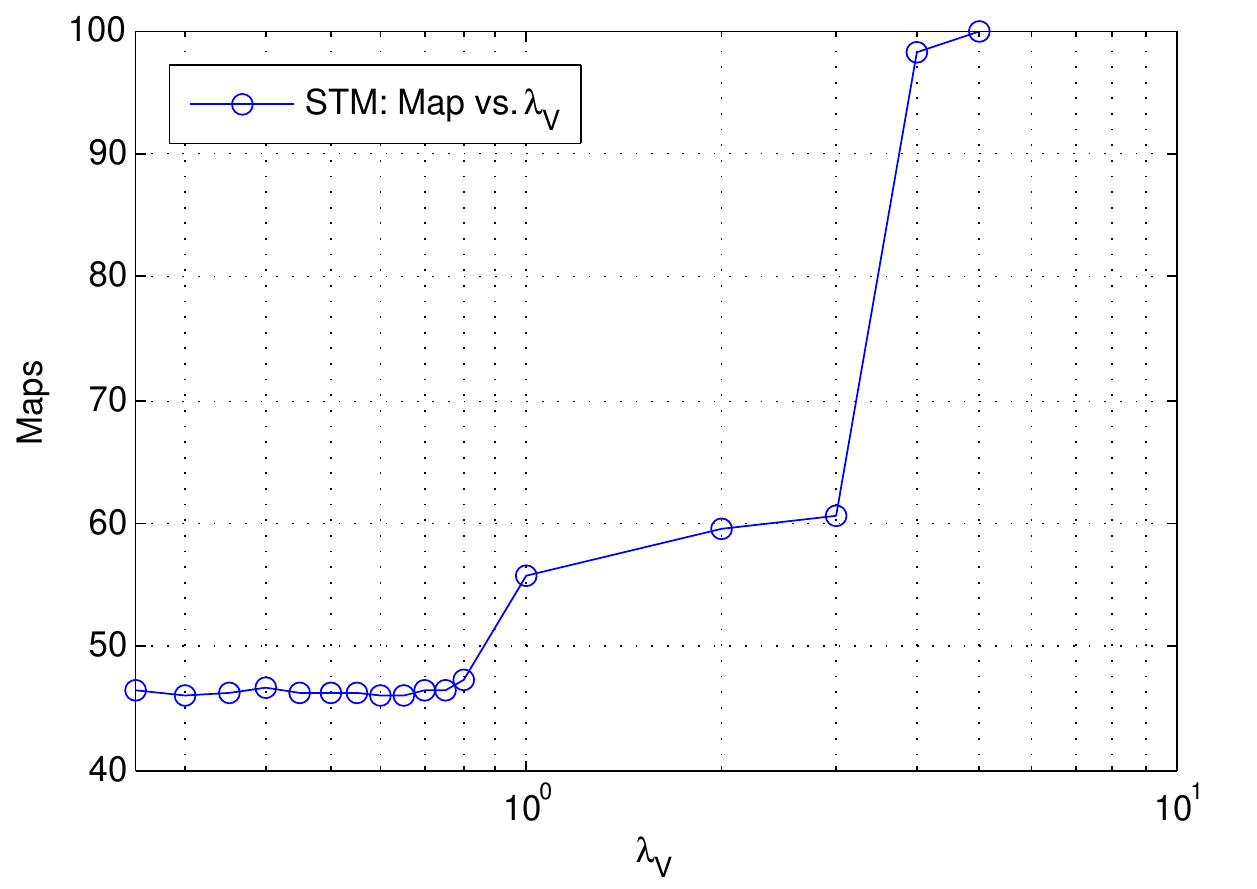}}
    \hspace{2.3in}
  \subfigure[mAPS vs. $\lambda_R$]{
    \label{fig:lambda_R}
    \includegraphics[width=0.95\columnwidth]{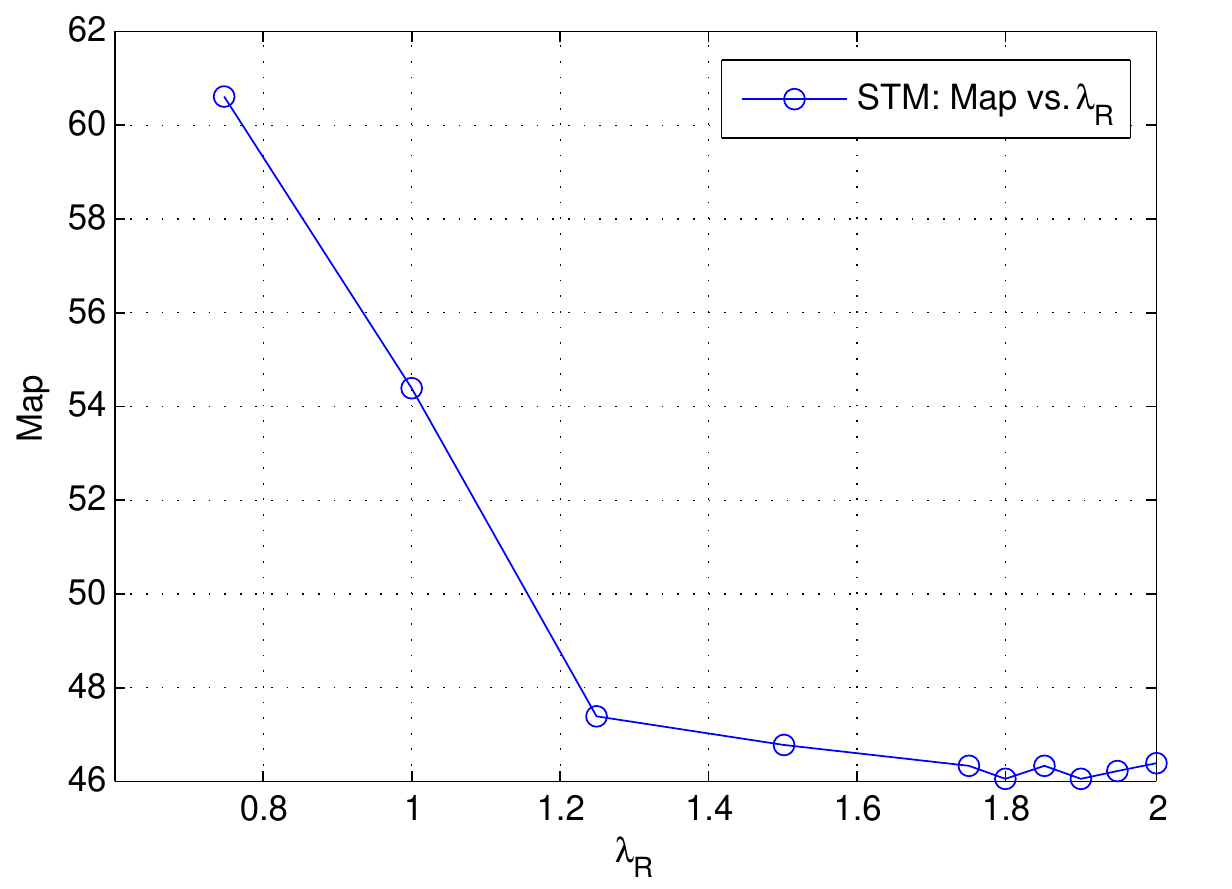}}
  \subfigure[mAPS vs. $K$]{
    \label{fig:K}
    \includegraphics[width=0.95\columnwidth]{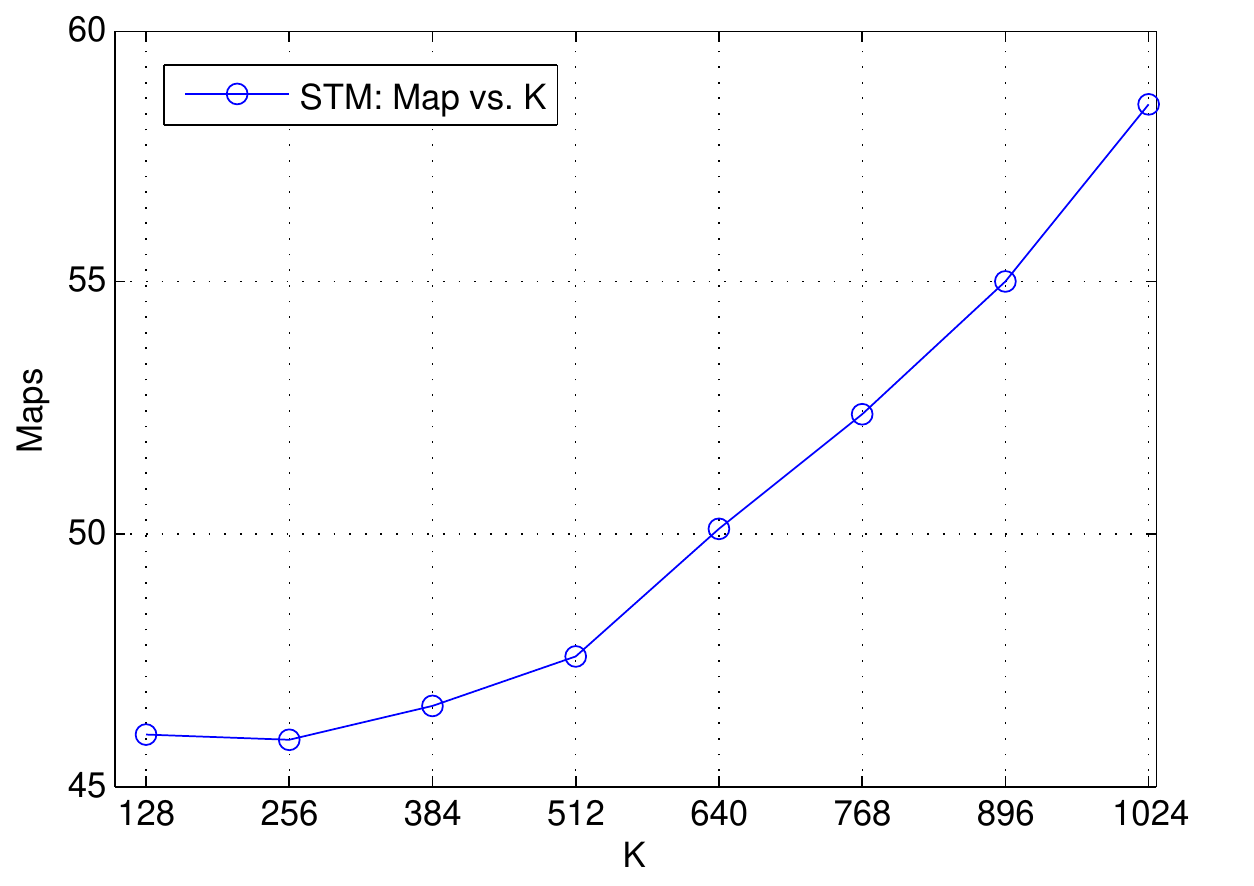}}
  \caption{mAPS on different parameter settings}
  \label{fig:parameter} 
\end{figure*}

We test the performance of {\em STM} by varying four parameters including $\lambda_R$, $\lambda_U$, $\lambda_V$, and the number of latent topics $K$ in {\em STM} as shown in Figure~\ref{fig:parameter} (a) - (d), respectively. For each test, we vary only one parameter and fix the other three.

According to these results, for the parameters $\lambda_U$ and $\lambda_V$ which control the constraints of sparsity in Eqn~\eqref{eqn:stms_obj}, the proposed method is very robust within the range $[0.1 - 0.8]$, and thus the {\em STM} is not sensitive to the selection of these two parameters.
When $\lambda_U$ and $\lambda_V$ become larger (in the experiments, when larger than $2.0$), the recommendation results are so sparse that the evaluation scores are worse.
In the following experiments, we use $\lambda_U = 0.35$ and $\lambda_V = 0.60$.

As for $\lambda_R$, which controls the balance between image content and user ratings, the results shown in Figure~\ref{fig:lambda_R} suggest that we should prefer the user rating history slightly by choosing the optimal $\lambda_R = 1.90$. Moreover, the performance is still not sensitive on the parameter $\lambda_R$ selection when $\lambda_R \geq 1.2$.

Finally, for the dimensionality of the latent topics space $K$ (or equivalently, the number of latent topics), the best interval is between $128$ and $256$.
This is reasonable because for the sparse coding, a typical dictionary size is 512 or 1024, while for the topic models, especially the CF, a feasible solution usually contains 100 to 200 topics.
Therefore, we choose $K = 256$ in our experiments which balances both visual features and latent topics.

\subsubsection{Sparsity}

Table~\ref{tab:sparsity} shows the sparsity of our obtained model using \emph{STM}. As suggested by these results, the user-profile and item-profile are rather sparse which coincides with our basic assumption in this paper. Moreover, this means each image covers $0.0084*256 = 2.15$ topics on average, while each user favorites $0.0112 * 256 = 2.87$ topics. This is reasonable as we discussed in Section 1, that the images are focusing on limited topics rather than ``thousands of words'' and an ordinary user prefers limited topics.

\begin{table}[h]
\centering
\caption{The sparsity of User-profile and item-profile}
\begin{tabular}{c|c|c}
\shline
                    &   User-profile $U$        &   Item-profile $V$\\
\hline
\textbf{Sparsity}   &   0.0112                  &   0.0084\\
\shline
\end{tabular}
\label{tab:sparsity}
\end{table}

To summarize, the proposed {\em STM} has the following properties:
\begin{enumerate}[1) ]
\item Quick Convergence:the algorithm will converge within 10 iterations;
\item Robustness on parameters: the algorithm is robust on different parameter choices, and the performance is not sensitive to the parameter values;
\item High sparsity: both the item and user profiles are highly sparse.
\end{enumerate}

\subsection{Experimental Comparison}


\begin{table}[h]
\centering
\caption{Performances of the proposed approaches compared with other baseline methods}
\label{tab:performance}
\begin{tabularx}{0.6\linewidth}{|X|X|}
\shline
\emph{method}       & \emph{mAPS}              \\
\hline
PMF \cite{pmf08}    & 61.99             \\
\hline
CTR-I               & 52.76                \\
\hline
SoRec\cite{ma08}    & 50.20     \\
\hline
\textbf{STM}        & \textbf{46.09}             \\
\hline
\textbf{SoSTM}      & \textbf{45.77}          \\
\shline
\end{tabularx}
\end{table}

\begin{figure}[htb]
\begin{center}
\centerline{\includegraphics[width=\columnwidth]{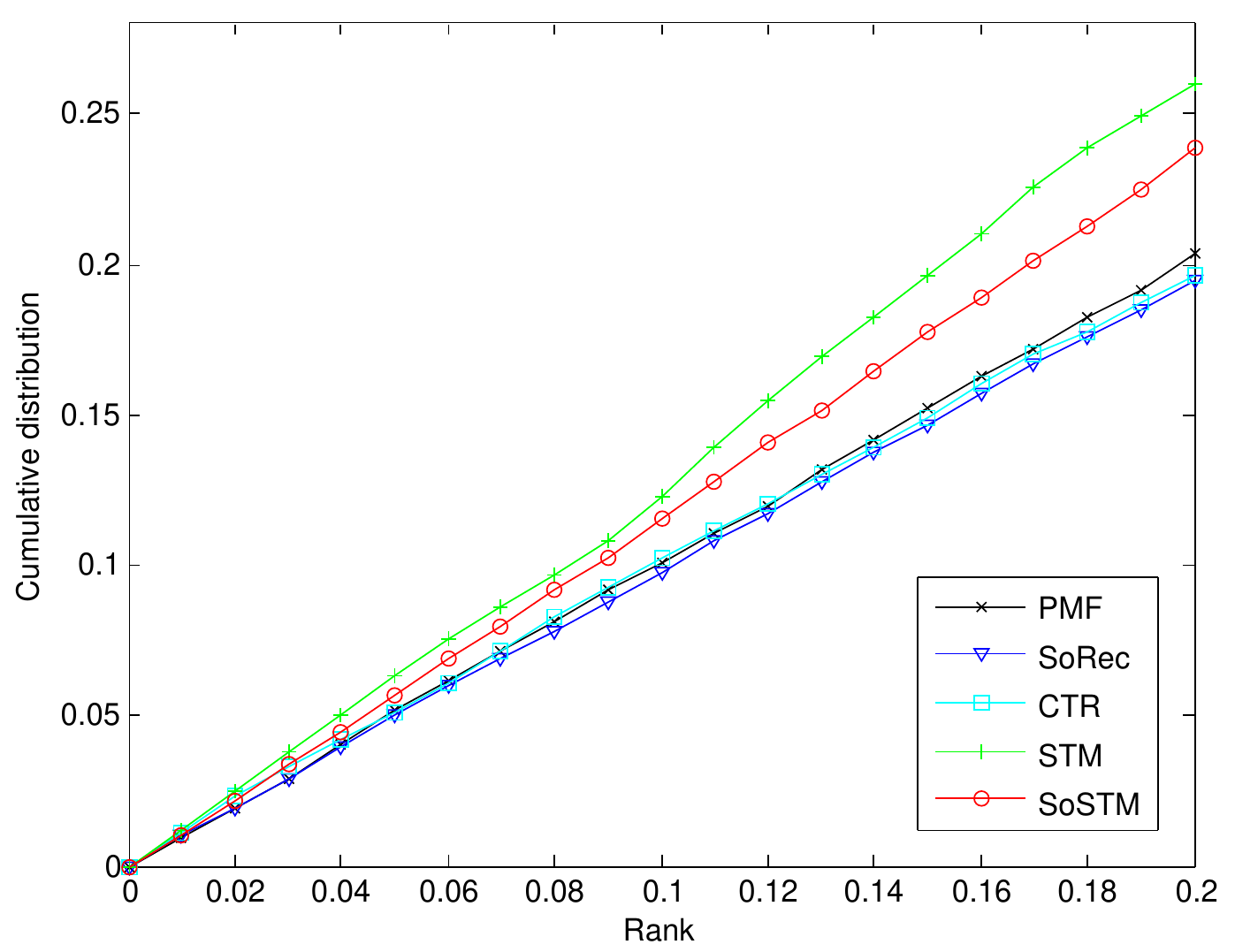}}
\caption{Cumulative percentage of
recommended image versus the ranking (percentile score)
with different algorithms on {\em FlickrUserFavor} data set.}
\label{fig:PR}
\end{center}
\end{figure}

We compare our algorithms, the {\em STM} and {\em SoSTM} namely, with several baseline approaches including stat-of-the-arts and newly proposed on the {\em FlickrUserFavor} data set in this section. The comparisons are shown in Table~\ref{tab:performance}.
We implement all the baseline methods on our data set. For {\em PMF} and {\em SoRec}, we use 30 dimensional latent features.
For {\em STM}, we adopt the parameters $\lambda_R = 1.90$, $\lambda_U = 0.35$, $\lambda_V = 0.60$; for {\em SoSTM}, $\lambda_S = 1.0$ and $\lambda_Z = 0.3$.
While for {\em CTR-I} in Eqn~\eqref{eqn:obj_ctr} we adopt the same setting as that for the {\em STM} approach.

As observed in the left column of Table~\ref{tab:performance}, without using the social hints information, the proposed {\em STM} algorithm achieve significant improvement over the stat-of-the-art {\em PMF}. This illustrates the promotion from content information in the recommendation task.
Furthermore, we also obtain an improvement from 52.76 to 46.09 over {\em CTR-I}, which suggests the effectiveness of introducing sparse coding in representing the visual content of images.

The right column of Table~\ref{tab:performance} shows the comparison between different algorithms by introducing the social hints. The social hints connect the related user by sharing their user profiles in the latent topic space, and hence improve the accuracy of recommendations as shown in Table~\ref{tab:performance} (note that the performances are improved significantly from {\em PMF} to {\em SoRec}). Moreover, our {\em SoSTM} also achieves a satisfying improvement by considering the social hints compared with original {\em STM}. Finally, among all the methods, the {\em SoSTM} achieves the best performance (mAPS = 45.77).

Figure \ref{fig:PR} plots a {\em P-PS} curve such that a point on the curve gives
the percentage of favored images from the testing data (Y-axis)
falling within a specific percentile score in the rank (X-axis).
Note that a higher  curve implies
that a larger percentage of the relevant images lie  within a
specific percentile score,  and this is desirable for our proposed {\em STM} and {\em SoSTM} approaches.


\subsection{Visualization of the Topics}

\begin{figure}[htb]
\centering
  \subfigure[]{
    \includegraphics[width=\columnwidth]{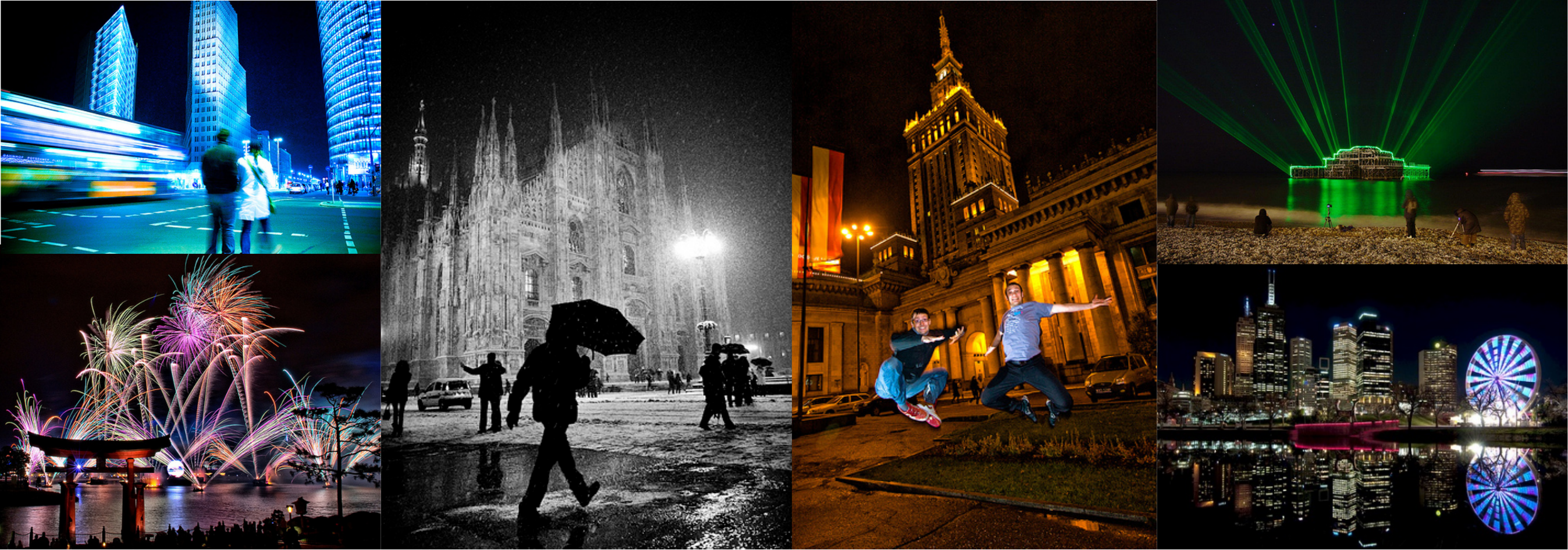}}
  \subfigure[]{
    \includegraphics[width=\columnwidth]{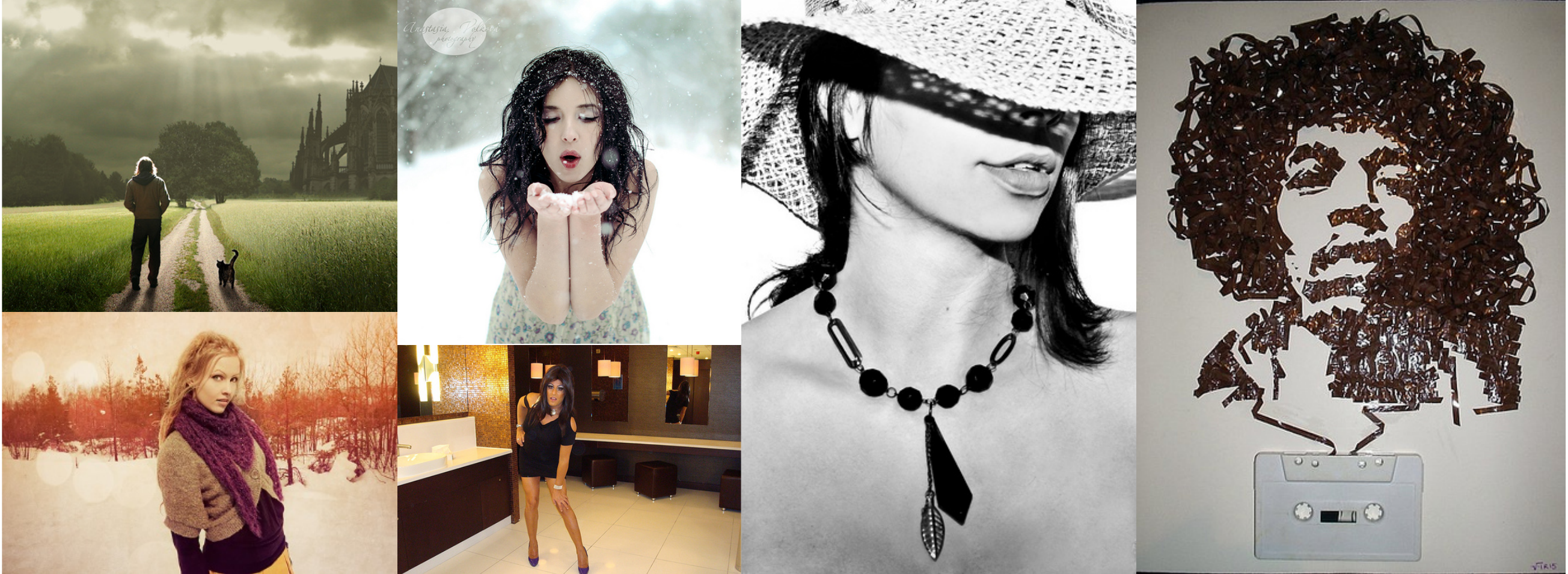}}
  \subfigure[]{
    \includegraphics[width=\columnwidth]{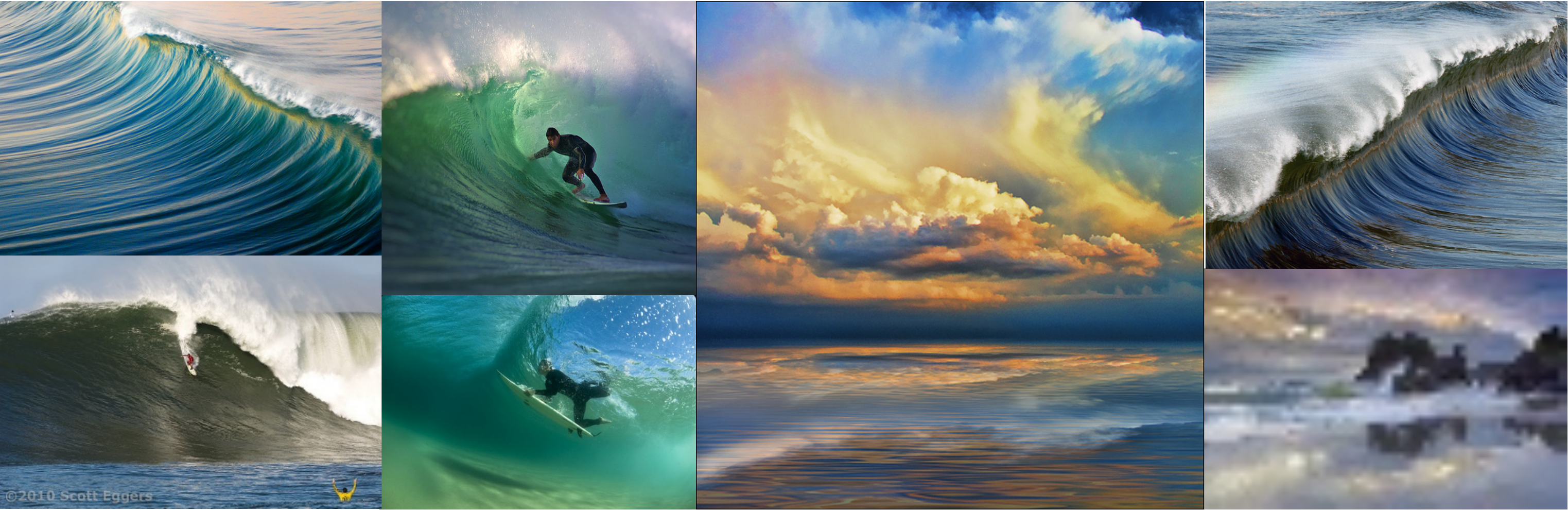}}
  \subfigure[]{
    \includegraphics[width=\columnwidth]{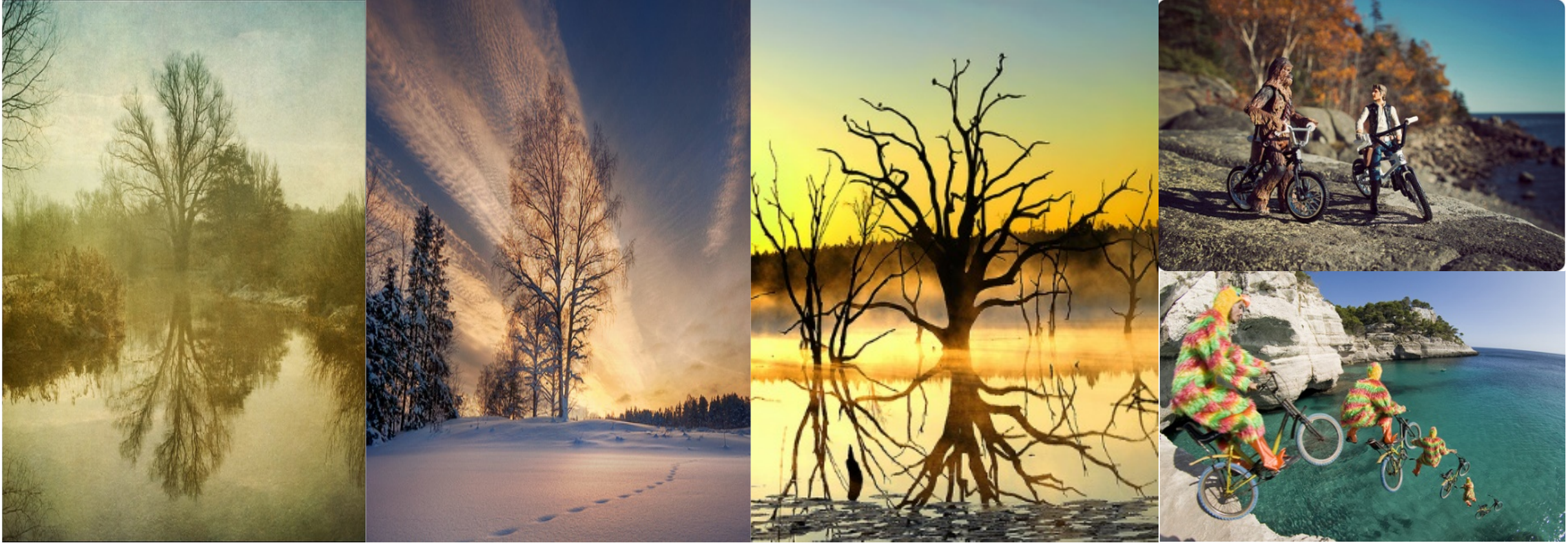}}
  \caption{Visualizations of latent topics learned by STM. Each row shows the top responded images for a specific topic.}
  \label{fig:topic_visualization} 
\end{figure}

By incorporating the visual content in our STM framework, we are able to understand the semantics of each topic in the latent space.
Figure~\ref{fig:topic_visualization} shows the visualizations of four topics learned by STM.
For each topic, the images with top responses on the corresponding bases are selected and shown.
As shown in the figure, the first topic (sub-figure a) is related with ``dark'', ``building'';
the second topic (sub-figure b) is more relevant with ``human'';
;the third topic (sub-figure c) is related with ``sea''; while the fourth one is related with ``tree'' and ``lake''.

The observations suggest that, our STM algorithm captures the most descriptive features within image contents,
and preserves certain level of semantics introduced by user ratings.

\subsection{Cold-Start for Unseen Images}

The cold-start problem is still a tough task far more from being solved. In this experiment, we are showing a potential attempt making use of the Sparse Topic Modeling algorithm.

In this experiment, we randomly leave out 20\% percent images as the unseen items, and use the entire or part of the rest images and their user ratings as the training data. In our experiments, we vary the percentage from 100\% (all the rest images) to 20\% in the training, and test the {\em STM} algorithm on the unseen images\footnote{Instead, if we varying the percentage of unseen images, as the unseen images become more and more, the expected mAPS will also increase. Finally, this will make the evaluation results not comparable.}.

\begin{figure}[htb]
\begin{center}
\centerline{\includegraphics[width=\columnwidth]{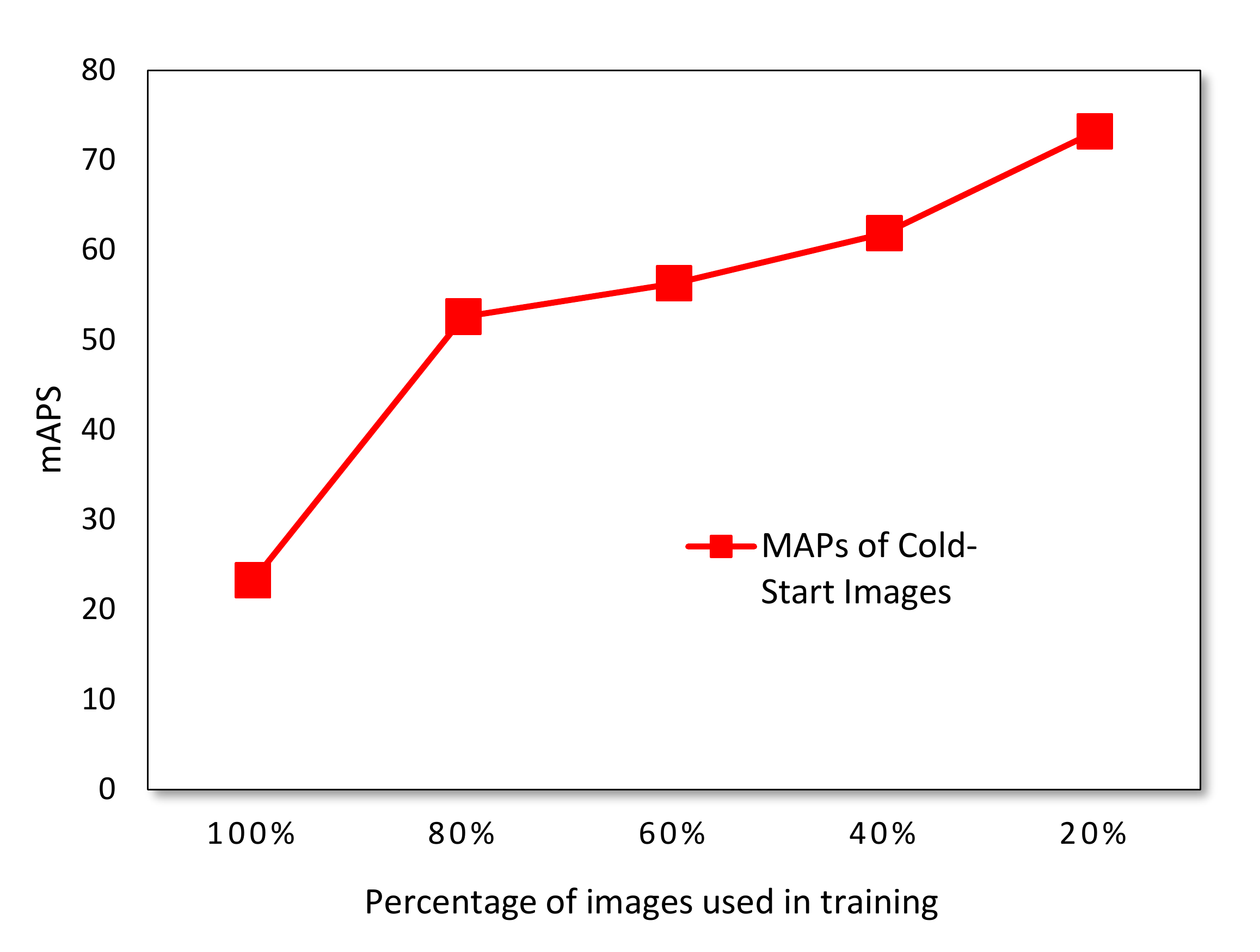}}
\caption{The performance when {\em STM} deals with cold-start images: 20\% images are left out as unseen images, and we use different percentage of the other images as the training data. As in the figure, the horizontal axis is the percentage of images used in training, the vertical axis is mAPS scores.}
\label{fig:cold-start}
\end{center}
\end{figure}

As shown in Figure~\ref{fig:cold-start}, the mAPS drops as fewer training images are used.
The {\em STM} can achieve satisfying results when all the ``seen'' images are utilized in the training, and
even when the percentage drops to only 20\%, it could still make effective recommendation to the users.
%

\section{Conclusion}
\label{sec:conclusion}

In this paper, we proposed a recommendation algorithm for images based on jointly learned user and image sparse representation in the latent topic space.
Different from the traditional collaborative filtering recommendation algorithms, our proposed method incorporates image content analysis,
therefore it is capable of making recommendations on cold start images. Compared with the other latent factor based methods, our approach not only achieves
superior performance, but also allows more compact storage of the user and image profiles, which is an extremely important advantage when dealing with real-world data.



\bibliographystyle{abbrv}
\bibliography{sigproc}  



%
%

\balancecolumns

\end{document}